\documentclass[twocolumn,showpacs,superscriptaddress]{revtex4-1}
\usepackage{mymacros}
\usepackage{graphicx}
\usepackage{epstopdf}
\pdfoutput=1
\usepackage{units}
\usepackage{multirow}
\usepackage{bigstrut}
\usepackage{overpic}
\usepackage{array}
\usepackage{amsfonts}
\usepackage{amsmath}
\usepackage{amssymb}
\usepackage{amsfonts}
\usepackage{verbatim}
\usepackage{footmisc}
\usepackage{booktabs}
\usepackage{color}
\usepackage{bm}
\usepackage{lipsum}
\newcommand{\ldc}{\Lambda_c^+}
\newcommand{\ldcb}{\bar\Lambda_c^-}
\newcommand{\ks}{K_S^0}
\newcommand{\Sigmaz}{\Sigma^0}
\newcommand{\piz}{\pi^0}
\newcommand{\half}{\frac{1}{2}}
\newcommand{\onehalf}{\frac{3}{2}}
\newcommand{\onehlf}{\frac{3}{2}}
\newcommand{\lpi}{\Lambda\pi^+}
\newcommand{\lbpi} {\bar\Lambda\pi^-}

\newcommand{\mbc}{M_{\rm{BC}}}
\newcommand{\tm} {\textrm}

\def \pp {\pi^0\pi^0}
\def \GeV {\,\rm{GeV}}
\def \MeV {\,\rm{MeV}}
\def \sig {$\sigma$}

\RequirePackage{lineno}
\begin{document}


\title{\large \bf Determination of the $\ldc$ spin via the reaction $\ee\to\ldc\ldcb$}


\author{
\begin{center}
M.~Ablikim$^{1}$, M.~N.~Achasov$^{10,c}$, P.~Adlarson$^{67}$, S. ~Ahmed$^{15}$, M.~Albrecht$^{4}$, R.~Aliberti$^{28}$, A.~Amoroso$^{66A,66C}$, M.~R.~An$^{32}$, Q.~An$^{63,49}$, X.~H.~Bai$^{57}$
% ...
\\
\vspace{0.2cm}
(BESIII Collaboration)\\
\vspace{0.2cm} {\it
$^{1}$ Institute of High Energy Physics, Beijing 100049, People's Republic of China\\
% ...
$^{4}$ Bochum Ruhr-University, D-44780 Bochum, Germany\\
% ...
$^{10}$ G.I. Budker Institute of Nuclear Physics SB RAS (BINP), Novosibirsk 630090, Russia\\
% ...
$^{15}$ Helmholtz Institute Mainz, Staudinger Weg 18, D-55099 Mainz, Germany\\
% ...
$^{28}$ Johannes Gutenberg University of Mainz, Johann-Joachim-Becher-Weg 45, D-55099 Mainz, Germany\\
% ...
$^{32}$ Liaoning Normal University, Dalian 116029, People's Republic of China\\
% ...
$^{49}$ State Key Laboratory of Particle Detection and Electronics, Beijing 100049, Hefei 230026, People's Republic of China\\
% ...
$^{57}$ University of Jinan, Jinan 250022, People's Republic of China\\
% ...
$^{63}$ University of Science and Technology of China, Hefei 230026, People's Republic of China\\
% ...
$^{66}$ University of Turin and INFN, (A)University of Turin, I-10125, Turin, Italy; (B)University of Eastern Piedmont, I-15121, Alessandria, Italy; (C)INFN, I-10125, Turin, Italy\\
$^{67}$ Uppsala University, Box 516, SE-75120 Uppsala, Sweden\\
% ...
$^{c}$ Also at the Novosibirsk State University, Novosibirsk, 630090, Russia\\
% ...
}\end{center}
\vspace{0.4cm}
}

\begin{abstract}
We report on a comparison of two possible $\ldc$ spin hypotheses, $J=\half$ and $\onehalf$, via the process $\ee\to\ldc\ldcb$, using the angular distributions of $\ldc$ decays into $p\ks$, $\Lambda\pip$, $\Sigmaz\pip$, and $\Sigmap\piz$. The data were recorded at $\sqrt s = 4.6$ GeV with the BESIII detector and correspond to an integrated luminosity of 587 pb$^{-1}$. The $\ldc$ spin is determined to be $J=\half$, with this value favored over the $\onehalf$ hypothesis with a significance
corresponding to more than 6 Gaussian standard deviations.
\end{abstract}

\maketitle

Since the discovery of the $\ldc$ particle more than 30 years ago \cite{Anjos:1987qe}, many other charmed baryons have been found and studied by experiments~\cite{pdg}. However, the $\ldc$ spin quantum number has not been determined conclusively until now. Unlike stable particles, whose spin can be measured with a dedicated detector, e.g. Stern-Gerlach setup, the spin of short-lived $\ldc$ can be only studied via its decays. Although the spin quantum number can be inferred from the phenomenological Regge trajectory \cite{chew,chenghy,chenhx,klempt,leebw,collins}, the establishment of the $\ldc$ spin needs a direct experimental measurement, making use of information on the angular distribution for the decayed particles. Thus a large size and clean data events are need in the analysis. The only previous investigation of this property was performed by the NA32 fixed-target experiment \cite{Jezabek}. The charmed baryon $\ldc$ was produced in the process $\pi^- \text{Cu} \to \ldc \bar DX $, where $X$ indicates the other particles produced from the interaction, and the decay $\ldc\to pK^-\pi^+$ was used to reconstruct the charmed baryon with 160 selected candidate events. The result was compatible with a spin-$1/2$ assignment, but was not conclusive due to the small sample size.

Currently, the spin of the $\ldc$ is inferred to be $\half$ from the naive quark model~\cite{Mann:1964}, in which charmed baryons are built from $udc$ quarks, and $\ldc$ is classified into the mixed-symmetric 20 multiplet with spin-1/2 assignment. Theoretically, the $\ldc$ system is suggested as a unique and excellent laboratory to study heavy quark symmetry and chiral symmetry of the $u,d$ light quarks \cite{chenghy,chenhx,klempt}. A large number of theoretical predictions on the $\ldc$ properties and decays are made based on the spin-1/2 assumption \cite{chenghy,chenhx,klempt}. Although the quark model works well for the ground states \cite{pdg}, experimental confirmation of the $\ldc$ spin is essential for testing the quark model spin assignment and theoretical predictions. Knowledge of the $\ldc$ spin is also important for measurement of its intrinsic properties, such as its anomalous magnetic moment \cite{Baryshevsky:2015zba}, magnetic dipole moments \cite{Bezshyyko:2017var} and electromagnetic dipole moments \cite{bagli,Baryshevsky:2019vou}. Moreover, its decays can be used as a spin polarimeter \cite{galanti} to determine the $c$-quark polarization at the Large Hadron Collider. Furthermore the $\ldc$ spin and polarization are intimately related to the understanding of other charmed baryon properties, {\it e.g.} the newly observed $\Xi_{cc}^{++}$ \cite{Aaij:2017ueg}, which decays into final states with $\ldc$.

In this Letter, an analysis of the $\ldc$ spin is performed via the process $\ee\to\ldc\ldcb$ at the center-of-mass (CM) energy $\sqrt{s} = 4.6\GeV$. The data accumulated with the  BESIII~\cite{Ablikim:2019hff} detector corresponds to an integrated luminosity of 587 pb$^{-1}$. We test the spin-1/2 and 3/2 hypotheses based on the angular distributions of the $\ldc$ decays into $p\ks$, $\Lambda\pi^+$, $\Sigma^0\pi^+$ and $\Sigma^+\pi^0$. The decays are studied by the single-tag method, \textit{i.e.} either the $\ldc$ or $\ldcb$ from $\ee\to\ldc\ldcb$ is reconstructed while the presence of its recoiled $\ldc$ or $\ldcb$ is inferred from kinematics. Throughout the Letter, the charged-conjugation modes are always implied, unless explicitly stated.

\begin{figure}[htbp]
\begin{center}
  \includegraphics[width=\linewidth]{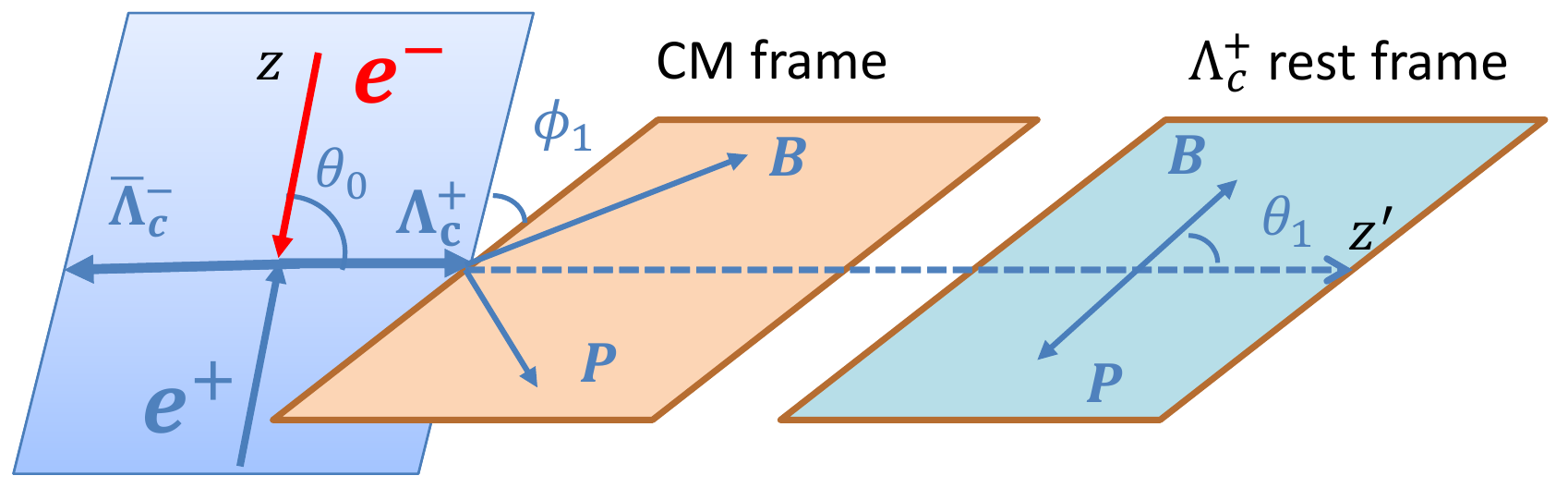}\\
  \caption{(color online) Definition of the helicity frame for $\ee\to\ldc\ldcb$, $\ldc\to BP$, where $B$ and $P$ denote a spin-$\half$ baryon and a meson, respectively.}
\label{fig1}
\end{center}
\end{figure}

The helicity formalism \cite{jacobi,chung} is applied in order to examine the implications of the $\ldc$ spin hypotheses for the joint angular distribution of the charmed baryon and its daughter particles. Figure \ref{fig1} shows the helicity frame for the $\ee\to\ldc\ldcb$ process. The helicity angle, $\theta_0$, is defined as the polar angle of the $\ldc$ in the $\ee$ CM system, with the $z$ axis pointing along the positron beam direction. For the $\ldc$ decay into a spin-$\half$ baryon ($B$) and a pseudoscalar meson ($P$), the $z'$ axis is defined along the direction of the $\ldc$, and $y'$ axis along $\hat z\times \hat z'$, and the $x'$ axis is determined by $\hat y'\times \hat z'$. The helicity angle $\phi_1$ is defined as the angle between the $\ldc$ production and decay planes and the helicity angle $\theta_1$ is the angle between the $B$ momentum in the $\ldc$ rest frame and the $z'$ axis.
The helicity angles for the subsequential baryon $B$ decays, $(\theta_i, \phi_i)$ with $i>2$, can be defined following the same procedure.

The two $\ldc$ spin hypotheses, $J=\half$ or ${3\over 2}$, are tested using  $\mathcal{W}^J$, which is the trace of the product of three matrices describing the joint angular distribution of the sequential decays:
\begin{equation}
  \mathcal{W}^J = \text{Tr}[\rho_J\cdot \mathcal T_J\cdot \mathcal T_B].
\end{equation}
Here $\rho_J$ is the spin density matrix for a $\ldc$ baryon produced in the process $\ee\to\ldc\ldcb$, $\mathcal{T}_J$ is a matrix describing the $\ldc$ decay to a baryon $B$ and a pseudoscalar meson $P$,
and the baryon $B$ sequentially decaying to the final states is described with a matrix $\mathcal{T}_{B}$.  The full formulas can be found in Refs \cite{chenh,angleLambdac,alphaLdc}. As an example, the $\ldc\to p \ks$, decay matrix, $\mathcal{T}_{B}$, reduces to the identity matrix, and the joint angular distribution is given by
\begin{eqnarray}
\mathcal{W}^{J=\half}(\theta_0,\theta_1,\phi_1) &\propto& 1+\alpha \cos^2\theta_0 + \mathcal{P_T} \sin\theta_1\sin\phi_1,\label{half-spin}\\
{\textrm{~with~}} \mathcal{P_T}& =& \alpha_{[p\ks]} \sqrt{1-\alpha^2}
   \cos\theta_0\sin\theta_0\sin\xi,\nonumber
\end{eqnarray}
where $\alpha$ is the angular-distribution parameter of the $\ldc$, $\alpha_{[p\ks]}$ the asymmetry parameter for the $\ldc\to p\ks$ weak decay and $\xi$ the relative phase between the two independent helicity amplitudes of the produced $\ldc$.

The joint angular distribution derived for the spin-${3\over 2}$ hypothesis for $\ldc\to p\ks$ is \cite{angleLambdac}
\begin{eqnarray}
\mathcal{W}^{J={3\over 2}}(\theta_0,\theta_1,\phi_1)  &\propto &
40 r^0_0 - 10 \sqrt{3} r^2_0(3\cos 2 \theta_1+1) \nonumber\\
&&\hspace{-2.5cm} -60 \left[r^2_1 \sin 2 \theta_1 \cos \phi_1 + r^2_2 \sin^2 \theta_1 \cos 2 \phi_1 \right]\nonumber\\
&&\hspace{-2.5cm}  + \sin \theta_1 \alpha_{[p\ks]} \big[ 8 \sqrt{15} r^1_{-1} \sin \phi_1 \nonumber\\
&&\hspace{-2.5cm}  + 90 r^3_{-2} \sin 2 \theta_1 \sin 2 \phi_1 \nonumber \\
&&\hspace{-2.5cm}  - 9 \sqrt{10} r^3_{-1} (5 \cos 2 \theta_1+3) \sin \phi_1 \big],
\end{eqnarray}
where the real multipole parameters, $r^L_M$, are defined in terms of the helicity amplitudes for charmed baryon pair production \cite{angleLambdac}.

The BESIII detector is an approximately cylindrically symmetric detector with $93\%$ coverage of the solid angle around the $\ee$ interaction point (IP). The components of the apparatus are a helium-based main drift chamber (MDC), a plastic time-of-flight (TOF) system, a 6240-cell CsI(Tl) crystal electromagnetic calorimeter (EMC), a superconducting solenoid providing a 1.0\,T magnetic field aligned with the beam axis, and a muon counter with resistive plate chambers as the active element.
The momentum resolution for charged tracks in the MDC is $0.5\%$ for a transverse momentum of $1\gevc$.
The photon energy resolution in the EMC is $2.5\%$ in the barrel region and $5.0\%$ in the end-cap region for $1\gev$ photons.
The combined information of the energy deposit in the MDC and the flight time measured by the TOF is used for particle identification (PID) of charged tracks. More details about the design and performance of the BESIII detector are given in Ref. \cite{Ablikim:2009aa}.

We use a large  Monte Carlo (MC) simulated sample of $\ee$ annihilations to understand background and to estimate the detection efficiencies.
The event generation is performed by the {\sc{kkmc}} generator \cite{Jadach:2000ir}, taking the beam-energy spread and initial-state radiation (ISR) into account.
Inclusive MC samples, consisting of generic $\lambdacp\lambdacm$ events, $D^{*}_{(s)}\bar{D}^{*}_{(s)}+X$ production \cite{Brambilla:2010cs}, ISR production of lower-lying charmonium(-like) $\psi$ states as well as continuum processes $\ee\to q\bar{q}~(q=u,d,s)$ are generated for a complete description of the background.
The decays are generated using {\sc{evtgen}}~\cite{Lange:2001uf} with the decay fractions from Ref.~\cite{pdg} as input.
The propagation through the detector and material interactions are simulated by using {\sc{geant4}}~\cite{ref:geant4}.

The  $\Lambda_c^+$ candidates are reconstructed from the $p\ks$, $\Lambda\pi^+$, $\Sigma^+\pi^0$, and $\Sigma^0\pi^+$ final states as done in Refs. \cite{lipr,alphaLdc}. The intermediate states, $\ks$, $\Lambda$, $\Sigma^+$, $\Sigma^0$ and $\pi^0$, are reconstructed from the $\pi^+\pi^-$, $p\pi^-$, $p\pi^0$, $\gamma\Lambda$ and $\gamma\gamma$ decays, respectively.

Charged tracks are required to originate from the interaction region, defined by a cylinder with a radius of $1\,\rm{cm}$ and a distance from the IP along the beam direction of $\pm10\,\rm{cm}$, except for those charged tracks coming from $\Lambda$ and $\ks$ decays. The polar angle $\theta$ of each track with respect to the beam direction is required to fulfill $|\cos\theta|<0.93$.
Protons are identified by requiring the PID likelihood $\Likelihood$ to satisfy $\Likelihood (p)>\Likelihood (K)$ and $\Likelihood (p)>\Likelihood (\pi)$, while charged pions are identified using $\Likelihood (\pi)>\Likelihood (K)$, except for those from $\Lambda$ and $\ks$ decays.

Clusters in the EMC with no associated charged tracks are identified as photon candidates if the energy deposit in the barrel region ($|\cos\theta|<0.80$) is larger than $25\,\rm{MeV}$, or if in the endcap region ($0.86<|\cos\theta|<0.92$) it is larger than $50\,\rm{MeV}$. To suppress background from electronic noise and coincidental EMC showers, the difference between the event start time and EMC signal is required to be smaller than $700\,\rm{ns}$. The $\pi^0$ candidates are reconstructed from photon pairs with an invariant mass, $M(\gamma\gamma)$, which satisfies $115<M(\gamma\gamma)<150\mevcc$. To improve the momentum resolution, a mass-constrained fit to the $\pi^0$ nominal mass is applied to the photon pairs and the resulting $\pi^0$ energy and momentum is used for the further analysis.

The $\Lambda(\ks)$ candidates are formed by combining the final states $p\pi^-(\pi^+\pi^-)$ with a displacement less than $20\,\rm{cm}$ from the IP along the electron beam direction. The $\chi^2$ of the vertex fit is required to be smaller than $100$ and the distance from the IP must be larger than twice the vertex resolution. The momenta of the daughter particles obtained from the fit are used in the further analysis. The charged pions associated with the $\Lambda$ and $\ks$ candidates are not subjected to any PID requirement, while proton PID is applied in order to improve the signal significance. To select $\ks$, $\Lambda$, $\Sigma^0$, and $\Sigma^+$ candidates, we require $487<M(\pi^+\pi^-)<511\,\mevcc$, $1111<M(p\pi^{-})<1121\,\mevcc$, $1179<M(\Lambda\gamma)<1203\mevcc$, and $1176<M(p \pi^0)<1200\mevcc$ , respectively. These requirements correspond to windows of approximately $\pm3$ standard deviations around the nominal masses.
In order to remove $p K_S^0,~K_S^0\to \pi^0\pi^0$ background in the $\Sigma^+\pi^0$ sample, the mass of the $\pi^0\pi^0$ pair is required to lie outside the range ($400, 550)\mevcc$.


\begin{figure}[t]
\centering
\includegraphics[width=\linewidth]{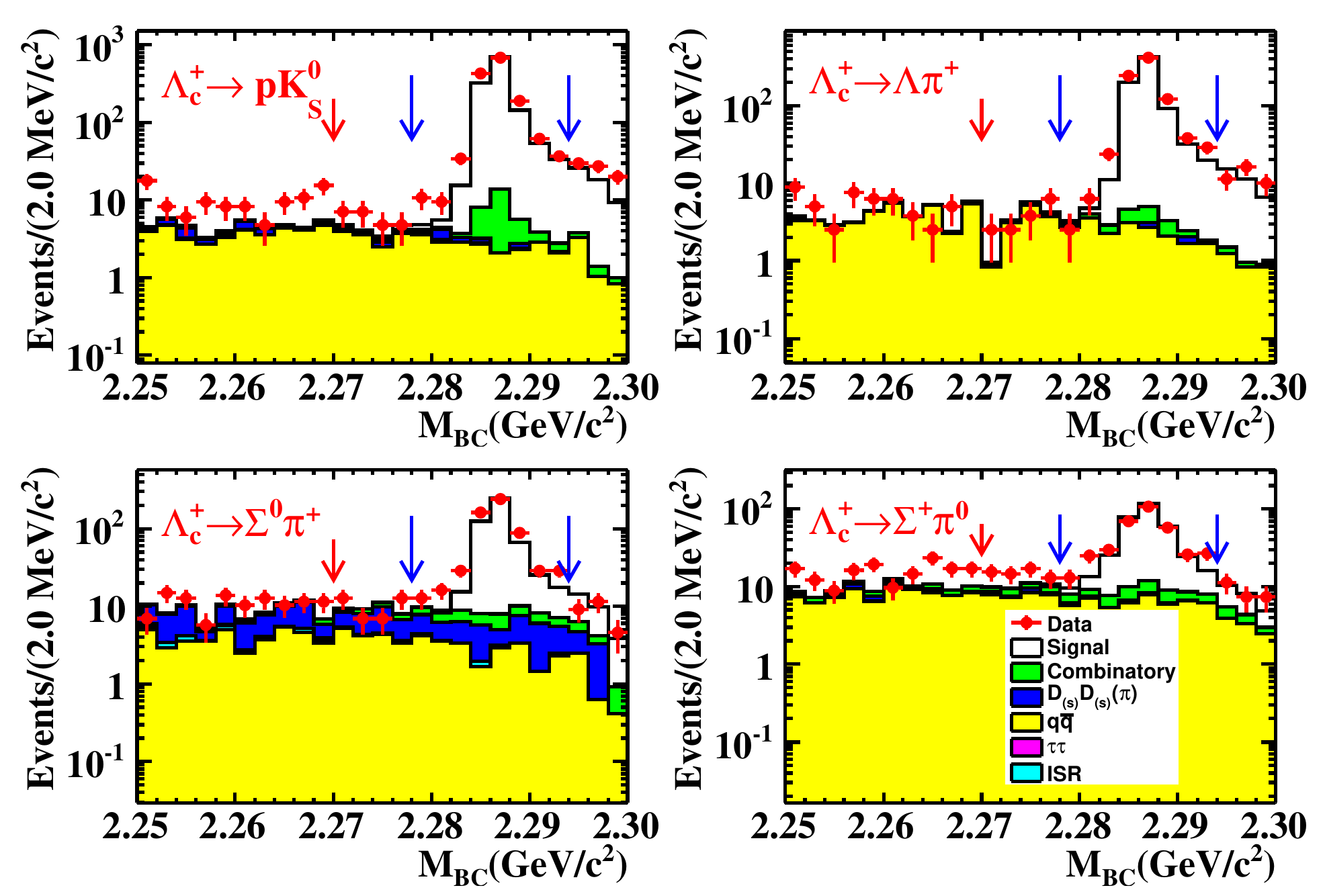}
\caption{ (color online) The $\mbc$ distributions for different decay modes.
Points with error bars represent the $\mbc$ distribution for the data, unfilled histograms for signal MC samples and shaded histograms for MC simulated background. The long vertical arrows indicate the $\ldc$ mass window, while the sideband region is to the left of the short arrow.}
\label{fig:mbc}
\end{figure}

The $\ldc$ candidates in each decay mode are selected by requiring the beam-constrained mass $\mbc\equiv\sqrt{E^2_{\rm beam}-p_{\Lambda_c^+}^2}$ to be within the range $(2.278, 2.294)\,\rm{GeV}/\emph{c}^2$, where $E_{\rm{beam}}$ is the beam energy and $p_{\Lambda_c^+}$ is the measured $\Lambda_c^+$ momentum in the CM system of the $e^+e^-$ collision. The numbers of reconstructed $\ldc$ candidates are 1227, 696, 614 and 412 for the $p\ks,~\Lambda\pi^+,~\Sigma^0\pi^+$ and $\Sigma^+\pi^0$ modes, respectively. If multiple candidates are found in a single event, we keep the one with the smallest energy difference $|\Delta E|$, where $\Delta E \equiv E_{\Lambda_c^+}-E_{\rm{beam}}$ and $E_{\Lambda_c^+}$ is the total measured energy of the $\Lambda_c^+$ candidate. To improve the signal purity, $\Delta E$ is required to be smaller than three times the resolution of energy difference distribution.  The $\mbc$ distributions of the different $\ldc$ decay modes are shown in Fig. \ref{fig:mbc}. The $\Lambda_c^+$ candidates appear as a peak at the nominal $\Lambda_c^+$ mass whereas the backgrounds, studied in inclusive MC samples, have smooth $\mbc$ distributions that are modeled with an Argus function~\cite{Argus}. The background level in the signal region can be estimated from sidebands, defined by $\mbc$ values within the range $(2.250, 2.270)\,\rm{GeV}/\emph{c}^2$. Table \ref{Tab:selecteEvt} lists the numbers of observed ($N^{\tm{obs}}$) and normalized background events ($N^{\tm{bg}}$) in the $\ldc$ signal region, where $N^{\tm{bg}}$ is estimated from the sideband.

\begin{table}[!h]
\begin{center}
\caption{Summary of observed ($N^\text{obs}$) and normalized background events ($N^\text{bg}$) in the $\ldc$ signal region, where $N^\text{bg}$ is estimated from the sideband. \label{Tab:selecteEvt}}
\begin{tabular}{ccc}
\hline\toprule\hline
Decay & $N^\text{obs}$ & $N^\text{bg}$ \\
\midrule\hline
$\ldc\to p\ks$         & 618      & 25.4  \\
$\ldcb\to \bar{p}\ks$ & 609      & 23.6 \\
$\ldc\to\lpi$         & 352       & 10.6 \\
$\ldcb\to \lbpi$      & 344       & 11.8 \\
$\ldc\to\Sigma^0\pi^+$       & 251   & 26.3 \\
$\ldcb\to \bar\Sigma^0\pi^-$ & 279   & 24.2 \\
$\ldc\to\Sigma^+\pi^0$       & 192   & 67.4 \\
$\ldcb\to \bar\Sigma^+\pi^0$ & 184   & 52.4 \\
\hline\bottomrule\hline
\end{tabular}
\end{center}
\end{table}

The $\ldc$ spin-$J$ hypotheses are tested using a likelihood function, which is defined for a given process $i$ as
\begin{equation}
\mathcal{L}^J_i(N^i)=\prod_{k=1}^{N^i}{1\over \mathcal{C}^i}\mathcal{W}^J(\theta_0^k,\theta_1^k,\phi_1^k,...,\theta_n^k,\phi_n^k),
\end{equation}
where $N^i$ is the number of events of $i$th decay mode defined in Table \ref{Tab:selecteEvt}, ($\theta_0^k,~\theta_1^k,~\phi_1^k,~...,~\theta_n^k,\phi_n^k$) are the helicity angles for the $k$-th event with $n$-step $\ldc$ decays, and $\mathcal{C}^i=\int \mathcal{W}^J(\theta_0,\theta_1,\phi_1,...,\theta_n^k,\phi_n^k) {\rm d}\cos\theta_0 \prod_{m=1}^{m=n}{\rm d}\cos\theta_m{\rm d}\phi_m$ is a normalization factor, calculated using a large phase-space MC sample.

The physics parameters are obtained by applying a simultaneous fit to the joint angular distribution of the selected events in the signal region. The background contributions are subtracted from the log-likelihood values using the weighted sideband events. The net log-likelihood for a given process $i$ is defined by
\begin{eqnarray}
\ln \mathcal{L}^J_i&=&
\ln\mathcal{L}^J_i(N_s^i)-\omega^{\tm{bg}}\ln\mathcal{L}^J_i(N_b^i),
\end{eqnarray}
where $N_s^i(N_b^i)$ is the number of selected data (background) events. The background weight, $\omega^{\tm{bg}}$, is the ratio between the number of background events in the signal region, and the number of sideband events. In estimating the background weight, its line shape in the fit is taken as an Argus function.

The {\sc{minuit}} \cite{minuit} package is used to minimize the objective function, $S=-\sum_i\ln \mathcal{L}^J_i$, in the simultaneous fit to the $p\ks$, $\Lambda\pip$, $\Sigmaz\pip$, and $\Sigmap\piz$ decay modes.
The decay asymmetry parameters for the spin-$\half$ hypothesis, \textit{e.g.} $\alpha_{[p\ks]}$ in Eq.~\eqref{half-spin} are constrained to the range $[-1,1]$ in the simultaneous fit. For the spin-${3\over 2}$ hypothesis, the asymmetry parameters in the fit are constrained to be in the physical region \cite{LeeYconstrain} for each mode, {\it i.e.} $-{1\over 3}\le \alpha_{[BP]}\le {1\over 3}$. The relative phase between the helicity amplitudes are fixed to the expected values near threshold \cite{angleLambdac}, whereas the moduli of the helicity amplitudes in $r^L_M$ are obtained from the fit.
The minimum log-likelihood, given by $-\sum_{i}\ln\mathcal{L}_i^{J}$, is determined to be $-45.18$ for the spin-$\half$ hypothesis and $-21.50$ for the spin-${3\over 2}$ hypothesis. Hence, our data favors the spin-$\half$ hypothesis.

\begin{figure}[htbp]
\begin{center}
\includegraphics[width=\linewidth]{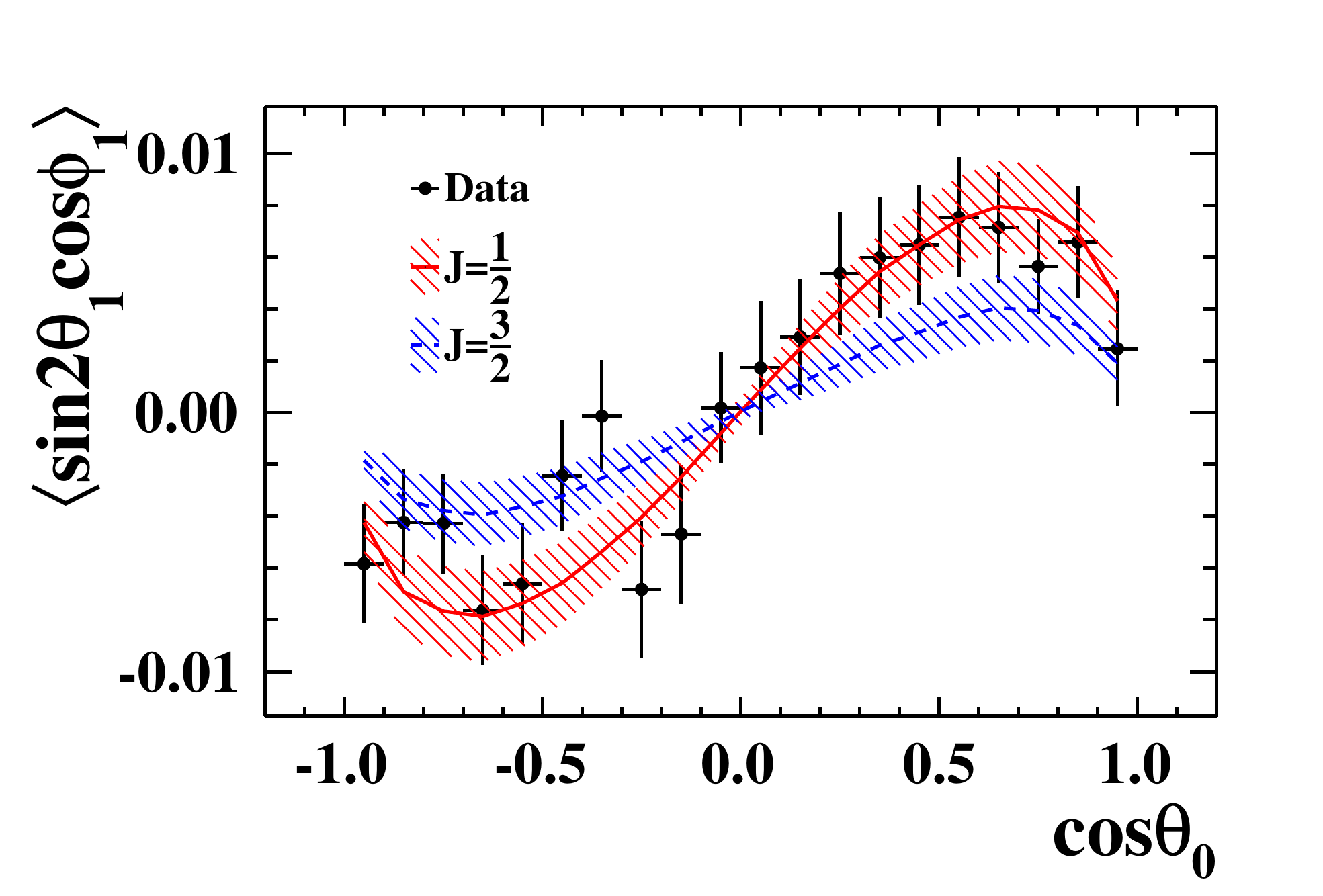}
\caption{(color online) The moments $\langle\sin2\theta_1\cos\phi_1\rangle$ as a function of $\cos\theta_0$. Points with error bars represent the combined data events from the tagged four $\ldc$ decay modes, where the background contribution has been subtracted. The red solid curve shows the fitted result of the spin-$\half$ hypothesis, whereas the blue dashed curve shows that of the spin-${3\over 2}$ hypothesis. The bands represent the fit uncertainty due to statistical uncertainties.}
\label{fitresultA}
\end{center}
\end{figure}

The distribution of estimates of the expectation value $\langle\sin2\theta_1\cos\phi_1\rangle$ moment, an average observed in each bin, is a useful observable to illustrate the different behaviour expected for the two hypotheses. Figure \ref{fitresultA} shows the first moment of the $\langle\sin2\theta_1\cos\phi_1\rangle$ distribution under the two spin hypotheses for the all-mode-combined events, and the projections of the two fits suggest that the data favor the spin-${1\over 2}$ to the spin-${3\over 2}$ hypothesis.

\begin{figure}[htbp]
\vspace{0.155cm}
\begin{center}
\includegraphics[width=0.98\linewidth]{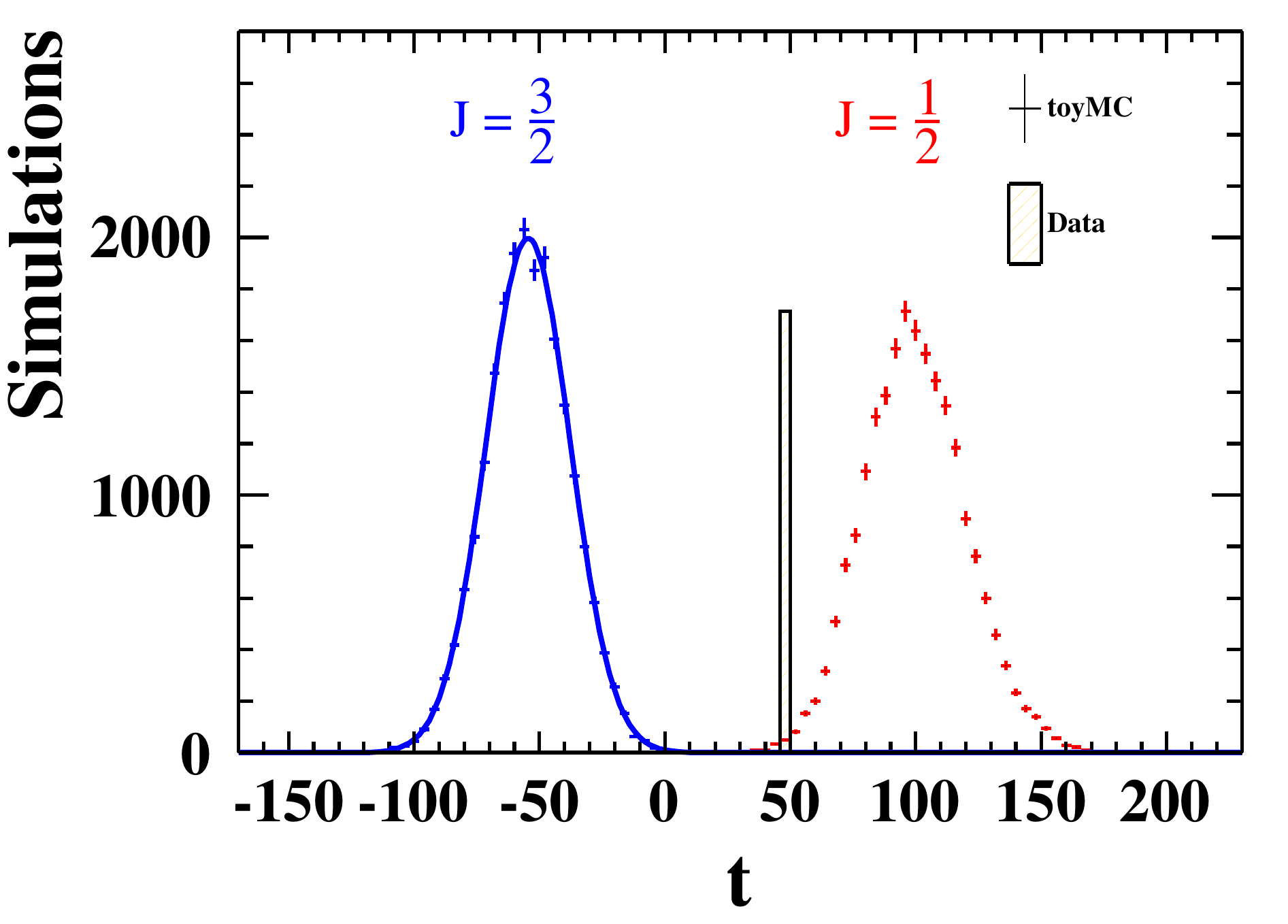}
\caption{(color online) Distributions of the test statistic $t\equiv-2\ln[\mathcal{L}^{J=3/2}/\mathcal{L}^{J=1/2}]$, for a series of MC simulations performed under the spin-${1\over 2}$ (right peak) and spin-${3\over 2}$ (left peak) hypotheses. The curve is the the Gaussian-fitted distribution to the left peak. The $t$ value
obtained from experimental data is indicated by the vertical bar.}
\label{toymc}
\end{center}
\end{figure}

In order to quantify the discriminating power of the test, we study the likelihood ratio distribution, $t\equiv-2\ln(\mathcal{L}^{J=3/2}/\mathcal{L}^{J=1/2})$, obtained from a series of MC simulations, following the method in Ref.~\cite{LHCb}. The MC sample for each hypothesis is generated according to its joint angular distribution, propagated through the detector model and subjected to the same event selection criteria as applied to the data events. Each MC subset has the same size as the data sample and is assumed to have the same amount of background. The test statistic $t$ distributions are shown in Fig.~\ref{toymc} for about 20,000 MC simulations. The simulations for the right peak ($t>0$) are performed under the $J={1\over 2}$ hypothesis, while those in the left peak ($t<0$) correspond to the $J={3\over 2}$ hypothesis. It is clear that the $t$-distributions of the two hypotheses are well separated, and can be discriminated between by setting an acceptance criterion of $t\ge 0$ for $J={1\over 2}$ and $t<0$ for $J={3\over 2}$. Since the $t$-value from the data fulfills $t\ge 0$, as shown in Fig.~\ref{toymc}, it is inconsistent with the spin-${3\over 2}$ hypothesis. Hence, our data favor the spin-$\half$ assignment.
The statistical significance for the spin-${3\over 2}$ over spin-${1\over2}$ hypothesis is estimated approximately with $(t_\text{data}-\langle t\rangle)/\sigma(t)$~\cite{LHCb}, where $\langle t\rangle$ and $\sigma(t)$ are the mean and standard deviation for the ensemble of MC simulations under the spin-${3\over 2}$ hypothesis with $t<0$. We find that the spin-${3\over 2}$ hypothesis can be rejected with a significance of $6.07\sigma$ in favour of the spin-$\half$ hypothesis.

\begin{table}[htbp]
\caption{Systematic uncertainties for estimating the significance of spin-$\half$  versus the spin-$\onehlf$ hypothesis determined with the toy MC method.\label{syserr}}
\begin{center}
\begin{tabular}{cc}
\hline\toprule\hline
Source                & Significance \\
\midrule
Nominal fit            & 6.07\sig \\
\hline
Tracks and PID         &  \\
Photon                 & 6.16\sig\\
$\pi^0$                & (combined)\\
$\Lambda$              &  \\
\hline
Sideband window        & (6.26,\,6.45)\sig \\
Signal window          & (5.92,\,6.07)\sig \\
$\Delta$E              & (5.77,\,6.39)\sig \\
$\omega^{\tm{bg}}$     & (6.37,\,6.41)\sig \\
$M_{\pi^0\pi^0}$ veto  & (6.06,\,6.30)\sig \\
\hline\bottomrule\hline
\end{tabular}
\end{center}
\end{table}

The significance to accept the spin-$\half$  hypothesis over the spin-$\onehlf$ hypothesis can be affected by the systematic sources listed in Table~\ref{syserr}. We estimate these systematic variations with the same MC method that was used for the likelihood ratio. The results are listed in Table~\ref{syserr}.

The efficiencies of the tracking and PID for charged tracks and their dependence on  transverse momentum and polar angle are measured using a control data sample $J/\psi\to p\bar p\pi^+\pi^-$ decays~\cite{pppipi}.
The uncertainties associated with the detection efficiency of the radiative photon in $\Sigma^0\to\gamma\Lambda$ decays are assessed with a control sample of $J/\psi\to~\pi^+\pi^-\pi^0,\pi^0\to~\gamma\gamma$ decays. The efficiency differences between data and MC simulations are  determined to be 0.5\% and 1.5\% in the barrel and endcap region, respectively \cite{photon_eff}. The difference between data and MC simulation of the $\pi^0$ reconstruction efficiency in the $\Lambda_c^+\to \Sigma^+\pi^0 $ decay, and its dependence on momentum, is studied using the processes $\psip\to\pi^0\pi^0J/\psi$ and  $\ee\to\pi^0\omega$ at $\sqrt s=3773\MeV$.
The $\Lambda$ reconstruction efficiency is studied as a function of momentum and polar angle in the reaction $\Lambda_c^+\to~\Lambda + X$ \cite{XiaoDong}. To take into account the correlations between the different sources of correction and uncertainty, we perform an overall weighting of MC events in the fit according to these efficiency corrections. We determine the significance of spin-$\half$ hypothesis to be $6.16\sigma$ with this approach.

The systematic uncertainties due to the event selection criteria of the $\Delta E$, signal and sideband events are estimated by varying their requirements by 1 MeV. The uncertainty due to the $M_{\pp}$ rejection criterion in the $\ldc\to\Sigmap\piz$ channel is checked with a tight and loose requirement, {\it ie.}, $M_{\pi^0\pi^0}\in [0.42,0.53]$\,GeV/$c^2$ and $[0.38,0.57]$\,GeV/$c^2$. The potential bias due to the sideband scale factor, $\omega^{\tm{bg}}$, is evaluated through varying the parameters by 1$\sigma$ for the Argus fit function. The ranges of significance estimation are given in Table~\ref{syserr}.
The resulting significance comparing the two hypothesis tests are found to be $6.07\sigma$ with a systematic boundary $(5.77\sim6.45)\sigma$, where the uncertainties correspond to the smallest and largest values listed in Table~\ref{syserr}.

In conclusion, we have compared the two spin hypotheses $\half$ and $\onehalf$ for the $\ldc$ baryon by
studying the process $\ee\to\ldc\ldcb$, using 587 pb$^{-1}$ of BESIII data collected at $\sqrt s=4.6\GeV$. The analysis considered the joint angular distribution of the production and decay modes $\ldc\to p\ks,~\Lambda\pi^+,~\Sigma^0\pi^+$ and $\Sigma^+\pi^0$. We found that the spin of $\half$ hypothesis is preferred over the $\onehalf$ with a significance of about $6\sigma$. Hence, we conclude the spin of the $\ldc$ baryon to be $\half$, consistent with the expectation of the naive quark model. Since the $\ldc$  is the lightest charmed baryon, this experimental determination of its spin is also a cornerstone in the extraction of the properties of heavier charmed and beauty baryons whose decay chains include this particle.

The BESIII collaboration thanks the staff of BEPCII and the IHEP computing center for their strong support.\\ This work is supported in part by National Key Basic Research Program of China under Contract No. 2015CB856700, 2020YFA0406300, 2020YFA0406400; National Natural Science Foundation of China (NSFC) under Contracts Nos. 11875262, 11835012, 11625523, 11635010, 11735014, 11822506, 11835012, 11935015, 11935016, 11935018, 11961141012; the Chinese Academy of Sciences (CAS) Large-Scale Scientific Facility Program; Joint Large-Scale Scientific Facility Funds of the NSFC and CAS under Contracts Nos. U2032110, U1732263, U1832207, U1832107; CAS Key Research Program of Frontier Sciences under Contracts Nos. QYZDJ-SSW-SLH003, QYZDJ-SSW-SLH040; 100 Talents Program of CAS; Fundamental Research Funds for the Central Universities; INPAC and Shanghai Key Laboratory for Particle Physics and Cosmology; ERC under Contract No. 758462; German Research Foundation DFG under Contracts Nos. Collaborative Research Center CRC 1044, FOR 2359; Istituto Nazionale di Fisica Nucleare, Italy; Ministry of Development of Turkey under Contract No. DPT2006K-120470; National Science and Technology fund; STFC (United Kingdom); The Knut and Alice Wallenberg Foundation (Sweden) under Contract No. 2016.0157; The Royal Society, UK under Contracts Nos. DH140054, DH160214; The Swedish Research Council; U. S. Department of Energy under Contracts Nos. DE-FG02-05ER41374, DE-SC-0012069; Olle Engkvist Foundation under Contract No. 200-0605.


\end{document}